# A Study of Multifractal Analysis in $^{16}$O-AgBr Collisions at 60A and 200A GeV

Nazeer Ahmad, Tufail Ahmad, Omveer Singh, Shakeel Ahmad

Department of Physics, Aligarh Muslim University, Aligarh, India
Email: nazeerahmadna@gmail.com

## Abstract

A multifractal analysis to study the multiparticle dynamics in 60A and 200A GeV, $^{16}$O-AgBr collisions has been performed in the pseudorapidity phase space. Multifractal moments, $G_q$, as the function of pseudorapidity bin size for different order of the moments, q, have been calculated. The power-law behaviour has been observed in the considered data sets. The variations of multifractal dimension, $D_q$, and the multifractal spectral function, $f(\alpha_q)$, with order of the moments, $q$, have been studied thoroughly. $D_q$ is found to decrease with increasing order of moments, $q$, indicating thereby a self-similar behaviour in the multiparticle production for the considered collisions. We have also found a concave downward curve of multifractal spectral function with maximum at $q = 0$.

## Keywords

Multifractality, Pseudorapidity, Fluctuation, Generalized Dimension, Scaling

## 1. Introduction

The study of event-by-event fluctuations in the pseudorapidity windows of decreasing bin width in multiparticle production at high energy has divulged self-similar properties as speculated by Bialas and Peschanski [1] [2] which is known as intermittency. The concept of self-similarity is closely related to the fractal theory which is a natural consequence of the cascading mechanism prevailing in the multiparticle production process. The power law behaviour, scaled factorial moments, indeed implies the existence of some kind of fractal pattern [2] in the dynamics of particles produced in their final state of reaction.

In order to investigate the cascading mechanism [3] of multiparticle production in the framework of the multifractal technique, a formalism has been de-

N. Ahmad *et al.*veloped by R.C. Hwa [4] [5] for a systematic study of the fractal properties and it also provides an effective means of defining non-uniform rapidity distribution of produced particles in nuclear collisions. The aim of this technique is to verify the authenticity of basic scaling properties, which occur in multifractal theories, when applied to particle production for the dynamical mechanism responsible for the hadronisation process in nucleus-nucleus (AA) collisions [6] [7] [8]. The explanation of cascading as self-similar process in analogy with geometrical objects such as fractals has been allowed by scaling law [9].

It is worth-mentioning that the issues of possible signals of Quark Gluon Plasma (QGP) formation in relativistic heavy-ion collisions are still on discussion stage. Fluctuations with different features are visualized to occur in different types of phase transitions. A study of multiplicity fluctuations in the final state of collisions might enable one to find out new observables linked to the characteristics of different types of phase transitions.

Some workers [10] [11] have proposed entropy based tools which are commonly used to describe the dynamics of complex systems. In the last few decades, it has been observed that Tsallis entropy and multifractal technique are found to be useful to characterize on large scale scaling behaviour. Furthermore, the Shannon, Reyni and Tsallis entropies [12] related formalisms are used as multifractal measures. However, the concept of divergence is being employed to quantify the departure between two distributions. It is one of the most used divergences related to the Tsallis entropy.

It is also interesting to mention that A. Deppman [13] describes a system which has a fractal structure in its thermodynamical functions called as thermofractals. The author has mentioned that its thermodynamics is more naturally described by the Tsallis statistics and a relation between fractal dimension and entropy index has been established. Furthermore, according to the Deppman such complex hadronic system exhibits thermofractal structure.

Although many attempts have been made to study the fractal properties, using $\mu p$, $p\bar{p}$ and $e^+e^-$ data [14] [15], using the method of multifractal moments, $G_q$, however AA collisions are less studied using this method. In the present paper, an attempt is made to investigate various interesting features of multifractality in $^{16}$O-AgBr collisions at 60A and 200A GeV. The Hwa's proposal which is based on the general theory of multifractals [16] [17] allows us to estimate mass exponent, $\tau_q$, generalized dimension, $D_q$ and multifractal spectral function $f(\alpha_q)$ for different order of the moments. These parameters serve as the measure of the multifractal structure in the considered data sets.

## 2. Theory

In order to examine the dependence of multifractal moments, $G_q$, on pseudorapidity, $\eta$, which is defined as; $\eta = -\ln\tan(\theta_s/2)$, $\theta_s$ is the space angle of a secondary particle with mean direction of primary, a given pseudorapidity range $\Delta\eta = \eta_{max} - \eta_{min}$ is divided into $M_0$ bins of width $\delta\eta = \Delta\eta/M_0$. A multifractal moment is defined as;

$$G_q = \sum_{j=1}^{M} p_j^q \quad (1)$$

where, $p_j = n_j/n$, such that $n(= n_1 + n_2 + \cdots + n_M)$. $M$ denotes the number of non-empty bins. $q$ is a real number and may has both positive or negative values. Once $G_q$ is calculated, its average over the entire sample is determined;

$$\langle G_q \rangle = \frac{1}{N_{ev}} \sum_{1}^{N_{ev}} G_q \quad (2)$$

where, $N_{ev}$ stands for the total number of events. If there is self-similarity in the production of particles, $G_q$ moments can be written in the form of a power law;

$$\langle G_q \rangle = (\delta \eta)^{\tau_q} \quad \text{for } \delta \eta \to 0 \quad (3)$$

where, $\tau_q$ stands for mass exponents. The linear dependence of $\ln \langle G_q \rangle$ on $\ln \delta \eta$ over all the windows is related as;

$$\tau_q = \lim_{\delta \eta \to \infty} \left( \Delta \ln \langle G_q \rangle / \Delta \ln \delta \eta \right) \quad (4)$$

The multifractal spectrum $f(\alpha_q)$ is related to the mass exponents $\tau_q$ and calculated from Legendre transform as:

$$\alpha_q = d\tau_q/dq, \quad (5)$$

$$f(\alpha_q) = q\alpha_q - \tau_q. \quad (6)$$

The properties of $f(\alpha_q)$ for multifractal behaviour are defined [4] [5] [16] [17] by;

$$\frac{df(\alpha_q)}{d\alpha_q} = q, \quad \frac{d^2 f(\alpha_q)}{d\alpha_q^2} < 0 \quad (7)$$

This downward concave form of curve of $f(\alpha_q)$ has the following characteristics for multifractals;

1) $f(\alpha_q)$ is downward concave
2) $f(\alpha_q = \alpha_0)$ is maximal
3) $f(\alpha_q) < f(\alpha_0)$, for $q \neq 0$

The behaviour of $f(\alpha_q)$ shows that $f(\alpha_q) = \alpha_q = 1$ for all $q$ as a special case if there are absolutely no fluctuations. The width of $f(\alpha_q)$ is a measure of the size of the fluctuations and the value $f(\alpha_0) < 1$ is a measure of the number of empty bins.

The generalized dimensions may be defined as;

$$D_q = \frac{\tau_q}{q-1} \quad (8)$$

Different types of dimensions are given as under:

1) $D_0 = f(\alpha_0)$; Capacity dimension: This shows how the data points of multifractal pattern fill the phase space domain

2) $D_1 = f(\alpha_1) = \alpha_1$; Entropy dimension: This is a measure of order-disorder

of the data points in the phase space domain under study. Larger the value higher the disorder.

3) $D_2 = 2\alpha_2 - f(\alpha_2)$; Correlation dimension: This quantifies the degree of clustering. Larger the value corresponds to higher level clustering.

## 3. Experimental Details

The data analyzed in the present paper were collected using two emulsion stacks exposed to Oxygen beams at 60A and 200A GeV at CERN, SPS (EMU01 Collaboration) [7] [18] [19] [20]. Along-the-track scanning method has been used to record the interactions in emulsion because of inherent high detection efficiency [7] [18] [21] [22]. In each event the polar and azimuthal angles of all the particles emitted in the interactions were measured with high magnification microscope. Tracks of the particles in the interactions were classified on the basis of their relative ionization, $g = I/I_0$ where $I_0$ and $I$ are respectively the ionizations of singly charged particle and charged particle to be identified. The tracks having relative ionization $g \geq 10$ are known as black tracks and their number in an interaction is denoted by $N_b$. This ionization cut corresponds to relative velocity $\beta < 0.3$. The grey tracks have relative ionization in the range $0.3 \leq g \leq 0.7$ and $N_g$ denotes their number in an event. However, relativistic charged particles, called shower tracks have relative ionization, $g < 1.4$ and this ionization cut corresponds to the relative velocity, $\beta > 0.7$.

Two data samples of 391 and 212 interactions of $^{16}$O with AgBr at 60A and 200A GeV respectively having $N_s \geq 10$ were used for analysis using standard emulsion criteria [23]. The pseudorapidity distribution for 60A and 200A GeV, $^{16}$O-AgBr interaction are plotted in **Figure 1**. In the figure, the red regions show the pseudorapidity ranges (1.6 - 3.6 ($\langle\eta\rangle \pm 1$) for 60A and 2.1 - 4.1 ($\langle\eta\rangle \pm 1$) for 200A GeV) which have been used in present analysis.

## 4. Results and Discussions

The variation of $\langle G_q \rangle$ as a function of $1/\delta\eta$ for $^{16}$O-AgBr at 60A and 200A GeV are plotted in **Figure 2** and **Figure 3**. The errors displayed in these figures represent the standard deviations from the mean values of $G_q$ moments.

From the figures, it may be noted that the moments with negative $q$ values saturate as $\delta\eta$ decreases whereas for positive q values, this shows linearity over a wide range of $\delta\eta$. This saturation could be due to decrease in number of particles as bin size is reduced. The variation of $\ln\langle G_q \rangle$ as a function of $-\ln\delta\eta$ also plotted in Figure 4. The data have been fitted using method of least squares for $-\ln\delta\eta$ 1. The linear rise of the multifractal moments with decreasing bin size of pseudorapidity shows a power-law behaviour in experimental data which is the indication of the self-similarity in the production mechanism of investigated reactions.

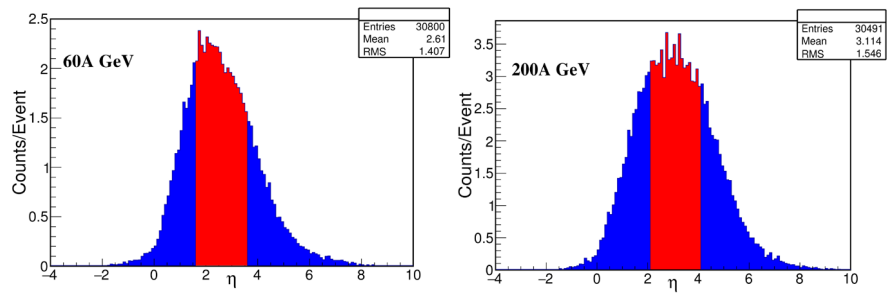

**Figure 1.** $\eta$ distribution for $^{16}$O-AgBr interactions at 60A and 200A GeV. The red regions show the pseudorapidity ranges of analysis.

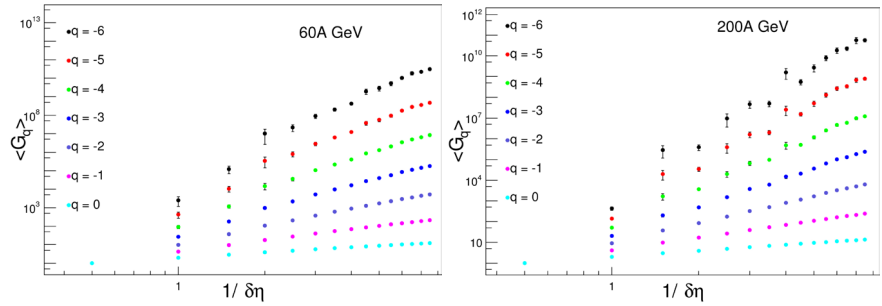

**Figure 2.** $\langle G_q \rangle$ as a function of $1/\delta\eta$ for $-6 \leq q \leq 0$.

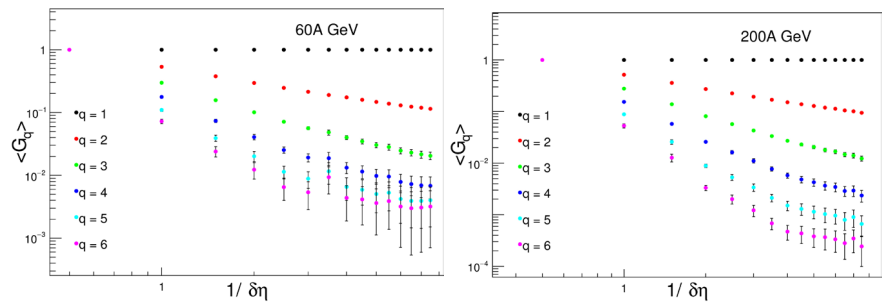

**Figure 3.** $\langle G_q \rangle$ as a function of $1/\delta\eta$ for $1 \leq q \leq 6$.

The mass exponents ($\tau_q$) have been calculated for the linear region of plots $\ln\langle G_q \rangle$ verses $-\ln\delta\eta$ and plotted as a function of *q* in Figure 5. It is observed from the figure that the values of $\tau_q$ increases linearly with increasing order of moments and independent of the collision energy.

The generalized dimensions, $D_q$, have been calculated using Equation (8) and plotted as a function of *q* in Figure 6 for both data sets. At both the energies $D_q$ decreases with increasing *q*, which shows the multifractal behaviour in multipartcle production. It may also be mentioned that for positive *q* values $D_q$ increases with increasing beam energy for same projectile, whereas for negative q values it seems to be independent of the projectile beam energy.

The multifractal spectrum, $f(\alpha_q)$, can also give an idea about the presence of multifractal behaviour in multipartcle production. We have calculated the multifractal spectrum $f(\alpha_q)$ and plotted in Figure 7. It is represented by a continuous concave downward curve with maximum at *q* = 0, $f(\alpha(0)) = D(0) = 1$ and the dotted line represents a common tangent at an angle of 45° at $\alpha_1 = f(\alpha_1)$.

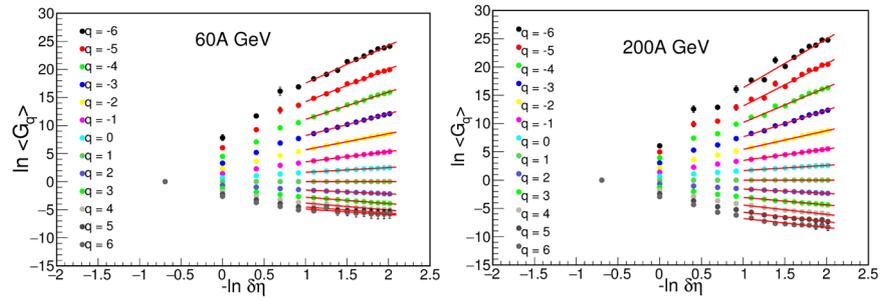

**Figure 4.** $\ln \langle G_q \rangle$ as a function of $-\ln \delta \eta$ for $-6 \leq q \leq +6$.

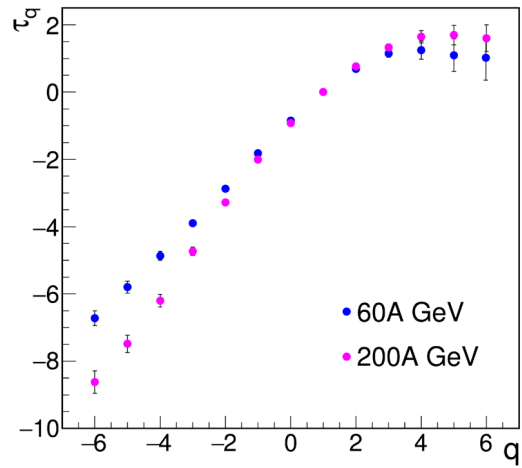

**Figure 5.** Variations of $\tau_q$ with $q$ for $^{16}$O-AgBr interactions at 60A and 200A GeV.

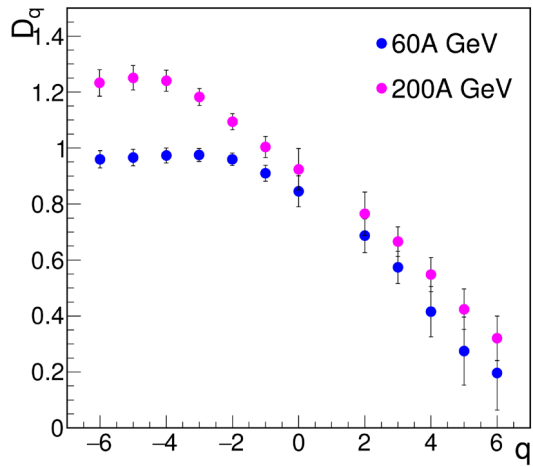

**Figure 6.** $D_q$ versus $q$ plots for 60A and 200A GeV, $^{16}$O-AgBr collisions.

These observations are in accordance with those reported earlier [24] [25] [26] [27]. However, all the spectra are wide enough to indicate the occurrence of multifractality in the multiparticle production process in 60A and 200A GeV $^{16}$O-AgBr collisions. It is also observed from **Figure 7** that the width of $f(\alpha)$

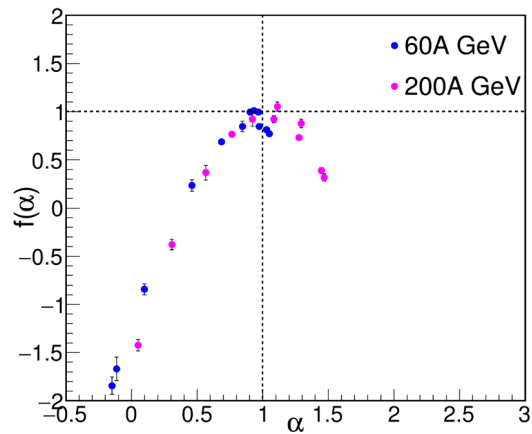

**Figure 7.** $f(\alpha_q)$ versus $\alpha_q$ plots for 60A and 200A GeV, $^{16}$O-AgBr collisions.

increases with increasing beam energy for the same projectile.

## 5. Conclusion

On the basis of the results discussed in the present paper, it is observed that, the moments having positive q values saturate with decrease in the bin size. Further, the variation of mass exponent, $\tau_q$, with the order of the moment, $q$, shows a power law behavior. As far as multifractality is concerned, a decreasing trend in the variation of multifractal dimension, $D_q$, with $q$ is observed. This indicates the presence of multifractality in the particle production process for the considered nuclear reactions. The observed behavior of multifractal spectral function $f(\alpha_q)$ in pseudorapidity space manifests self similarity in the mechanism of multiparticle production

## Acknowledgements

We are grateful to Professor Anju Bhasin, Jammu University, Jammu, INDIA for providing the experimental data used in the present analysis.